\documentclass[pre, notitlepage, nofootinbib, twocolumn, unsortedaddress, final]{revtex4-1}

\usepackage{xcolor}
\usepackage{graphics}
\usepackage{epsfig}

\usepackage{amsfonts, amsmath, amssymb}
\usepackage{enumerate}

\renewcommand{\Re}{\ensuremath{\mathrm{Re}}}
\renewcommand{\Im}{\ensuremath{\mathrm{Im}}}

\begin{document}

\title{High-performance solution of the transport problem\\
in a graphene armchair structure with a generic potential}

\author{Demetrio Logoteta} 
\altaffiliation[Present address: ]{IMEP-LAHC, Grenoble INP-Minatec, 3 Parvis Louis N\'eel, 38016 Grenoble Cedex 1, France.}

\author{Paolo Marconcini} 
\affiliation{Dipartimento di Ingegneria dell'Informazione, Universit\`a di Pisa, Via Caruso 16, I-56122 Pisa, Italy.}

\author{Claudio Bonati} 
\affiliation{Dipartimento di Fisica, Universit\`a di Pisa and INFN, Largo Pontecorvo 3, I-56127 Pisa, Italy.}

\author{Maurizio Fagotti} 
\affiliation{The Rudolf Peierls Centre for Theoretical Physics, University of Oxford - Oxford OX1 3NP, UK.}

\author{Massimo Macucci} \email[Corresponding author: ]{m.macucci@mercurio.iet.unipi.it}
\affiliation{Dipartimento di Ingegneria dell'Informazione, Universit\`a di Pisa, Via Caruso 16, I-56122 Pisa, Italy.}

\begin{abstract}
  
We propose an efficient numerical method to study the transport properties of armchair graphene ribbons 
in the presence of a generic external potential.  
The method is based on a continuum envelope-function description with physical boundary conditions. 
The envelope functions are computed in the reciprocal space, and the transmission is then obtained with a 
recursive scattering matrix approach.
This allows a significant reduction of the computational time with respect to finite difference simulations.

\end{abstract}                                                                 

\maketitle

\section{Introduction}

Graphene, a two-dimensional hexagonal lattice of carbon atoms isolated
in 2004 by Geim and Novoselov~\cite{geim2004}, represents a very interesting material
in several fields of science and technology. Since its low-energy
properties can be described by a Dirac equation, it is considered to be an
ideal test bed for the investigation of relativistic effects at non-relativistic
velocities~\cite{katsnelson2007}. 
Moreover, its unique properties make it suitable for applications in many different
fields of technology~\cite{geim2009}. Its atomic thickness and high room-temperature mobility, 
for example, make it a candidate material for postsilicon electronics. 

For applications in digital electronics the presence of a sufficiently large energy gap is 
fundamental~\cite{iannaccone2009}. A gap can be induced 
by the lateral confinement 
in narrow ribbons with a transverse 
size of a few nanometers or of tens of nanometers, which can be efficiently
modeled with atomistic techniques, such as tight-binding approaches. 
Fabrication of such nanowires is, however, very challenging, and therefore also 
alternative and/or complementary approaches to open up a gap are being 
pursued, such as the usage of bilayer graphene~\cite{mccann2006,ohta2006,
castro2007,zhang2009}, of chemical 
functionalization~\cite{ouyang2008,echtermeyer2009,elias2009,boukhvalov2009},
and of doping~\cite{acs2012,iwceroche,tang2012,wei2009}.
Large graphene devices (with a size of several hundreds 
of nanometers or of microns) can, however, be convenient 
(or mandatory) in radio-frequency or sensor applications, 
which do not necessarily require 
an energy gap~\cite{novoselov2012,schwierz2010,mina2013,hill2011,bonaccorso2010,xia2009}.  
In these cases an atomistic analysis is not feasible, since it is numerically 
too expensive, and more approximate techniques (such as envelope-function descriptions) 
have to be adopted. The availability of techniques suitable for the simulation of 
large-area graphene structures is also essential for a direct comparison with 
transport and noise measurements
performed on micron-sized flakes (see for example Refs.~\cite{cooper2012,peres2010,danneau2008,balandin2013}). 

Numerical efficiency is particularly important for transport calculations, 
which are usually self-consistently coupled with the solution of the Poisson 
equation, and  therefore 
typically need hundreds or thousands of iterations to reach 
global convergence. 
As a consequence, numerical performance can often make 
the difference between a feasible 
calculation and a computationally  impossible task. For this reason, 
it is important to develop reliable algorithms that are more 
approximate but more efficient than the atomistic ones.

A continuum approach has been often used for the numerical simulation
of transport in ribbons made up of transverse regions
with constant potential~\cite{wurm2009,li2009,zhou2010}, for which
analytical expressions for the wave function in each region are available
\cite{castroneto2009,brey2006,kp2011}.
However, for a generic potential, the envelope function equation (which
is a Dirac equation) has to be solved numerically, and 
the relativistic dispersion relation introduces some complications
in the standard discretization schemes.

In order to avoid these difficulties, the first numerical studies~\cite{Bardarson,Nomura}
adopted a momentum space regularization of the Dirac equation.
For ribbons with a large aspect ratio, such as 
those studied in Refs.~\cite{Bardarson,Nomura},
the boundary conditions are expected to be largely uninfluential and thus periodic
boundary conditions were used. Tworzid{\l}o {\it et al.}~\cite{tworzydlo2008} 
later performed a real-space transport analysis of large aspect ratio graphene 
ribbons by adopting the Stacey discretization scheme~\cite{stacey1982, 
bender1983}. 
More recently, Hern\'andez {\it et al.}~\cite{hernandez2012} studied
the transport properties of zigzag and armchair ribbons in the direct space, 
with physical boundary conditions.  For the zigzag ribbon, in the transverse direction
they adopted the discretization by Susskind~\cite{susskind1977}
and in the longitudinal direction the discretization by Stacey; 
for the armchair ribbon they used the Stacey discretization in both directions.
Snyman {\it et al.}~\cite{snyman2008} previously adopted the alternative
method of mapping the Dirac equation onto a Chalker-Coddington network 
model~\cite{chalker1988},
which in the past was used to describe percolation in disordered samples
in the quantum Hall effect~\cite{ho1996,connolly2012,logoteta2012}.

After the seminal contributions by Bardarson \textit{et al.}~\cite{Bardarson} and by 
Nomura \textit{et al.}~\cite{Nomura} the attention of the graphene community 
mainly focused on the finite difference methods, for which 
the physical boundary conditions can be easily 
implemented~\cite{CCnote}. In this paper we will show how to extend 
the reciprocal space algorithm to the case of armchair boundaries, and we will
present compelling evidence that this method is numerically much more efficient than the finite difference schemes.

For the solution of the transport problem we will adopt a
scattering-matrix approach. The ribbon is partitioned into
a series of thin slices in the direction along the current flow
(which we will refer to as the longitudinal direction).
In each slice the potential is approximated with a longitudinally constant function,
and the Dirac eigenvalue problem is solved 
in the Fourier transformed space. The conductance of the whole structure is then 
evaluated applying a mode-matching procedure at the interfaces between adjacent slices.

The paper is organized as follows. 
In Sec.~\ref{s:problem} we introduce the $\vec k\cdot \vec p$ approximation
for graphene ribbons and present the equations that describe the 
transverse slices of the device. In order to perform a comparison with
the approach we have decided to adopt, in 
Sec.~\ref{s:direct} we outline a few finite-difference techniques
that could be employed to solve the envelope function equation
in the direct space. In Sec.~\ref{s:reformulation} we 
introduce a mapping of the armchair problem into one with periodic boundary conditions
and in Sec.~\ref{s:fourier} we provide a detailed discussion of the reciprocal space technique 
used in Ref.~\cite{PT}, where numerical precision was crucial.
In Sec.~\ref{s:comparison} we compare the numerical efficiency of the 
methods in the direct and reciprocal space.
Section \ref{s:transport} is devoted to the solution of the two-dimensional 
transport problem for an armchair graphene ribbon in the realistic situation 
of a potential that varies also in the longitudinal direction.

\section{The numerical problem}\label{s:problem}

The wave function of graphene can be approximated by means of a linear 
combination of the $2 p^z$ orbitals of the carbon atoms of its two inequivalent  
sublattices (see, e.g., Refs.~\cite{castroneto2009,ando2005,kp2011}). 
Using the subscripts $A$ and $B$ 
to distinguish the quantities associated with different sublattices, the wave 
function is written as
\begin{equation} \label{total}
\psi (\vec{r}\,)=\sum_{\vec{R}_A}\psi_A (\vec{R}_A)\varphi(\vec{r} -\vec{R}_A)+
\sum_{\vec{R}_B} i\,\psi_B (\vec{R}_B)\varphi(\vec{r} -\vec{R}_B) \, ,
\end{equation}
where the sums are over the atom positions, and $\varphi(\vec r\,)$ is the $2 p^z$ orbital.

We consider an armchair ribbon with $N_D$ dimer lines
of carbon atoms across its width. The distance between the opposite
edges  is $W=(N_D-1)\,a/2$, where
$a=\sqrt{3}a_{C-C}$ is the graphene lattice constant and $a_{C-C}\simeq 0.142$~nm is the distance
between nearest-neighbor atoms. We denote by $x$ and $y$ the longitudinal 
and transverse directions, respectively; the Dirac points $\vec K$ and 
$\vec K^\prime$ can be expressed as $\vec{K}=-K \hat{y}$ and $\vec{K'}=K 
\hat{y}$, with $K=4 \pi/(3 a)$. 

For large enough $N_D$ and a potential that is slowly varying on the lattice 
scale, atomistic details can be disregarded and the low-energy physics can be 
described by means of a $\vec k\cdot\vec p$ approximation. This
is achieved in practice by promoting the coefficients 
$\psi_A(\vec R_A)$ and $\psi_B(\vec R_B)$ in \eqref{total}
to continuous functions of the position ${\vec r}$ 
and by writing the functions $\psi_A (\vec{r})$
and $\psi_B(\vec{r})$ in terms of four envelope functions $F$
as (see, e.g., Refs.~\cite{castroneto2009,ando2005,kp2011})
\begin{equation}\label{psia}
\begin{aligned}
\psi_{\beta} (\vec r) = 
e^{i \vec K \cdot \vec r} F^{\vec K}_{\beta} (\vec r)-i\, e^{i \vec K' \cdot \vec r} F^{\vec K'}_{\beta} (\vec r)\, ,
\end{aligned}
\end{equation}
where $\beta=A,B$.

It can be shown (see, e.g., Refs.~\cite{castroneto2009,ando2005,kp2011})
that the envelope functions satisfy the massless Dirac equation
\begin{equation}\label{dirac0}
\begin{aligned}
\displaystyle 
-i\gamma(\partial_x\sigma_x+\partial_y\sigma_y)\vec{F}^{\vec{K}} &= E\vec{F}^{\vec{K}} \\
\displaystyle 
-i\gamma(\partial_x\sigma_x-\partial_y\sigma_y)\vec{F}^{\vec{K'}} &= E\vec{F}^{\vec{K'}}\,,
\end{aligned}
\end{equation}
where
\begin{equation} 
\vec{F}^{\vec{\alpha}}=\left[\begin{array}{c} F_A^{\vec{\alpha}} (\vec{r})\\ F_B^{\vec{\alpha}} (\vec{r})\end{array}\right]\, 
\end{equation}
(with $\vec{\alpha}=\vec{K},\vec{K'}$), $\partial_x=\partial/\partial x$, $\partial_y=\partial/\partial y$,
$\sigma_x$, $\sigma_y$ are Pauli matrices, $E$ is the total 
energy of a particle in the ribbon, and $\gamma=(\sqrt{3}/2)\gamma_0 a\equiv v_F \hbar$. 
The constant $\gamma_0\simeq 2.7$~eV is the modulus of the transfer integral between nearest-neighbor carbon 
atoms, $v_F$ is the Fermi velocity of graphene, and $\hbar$ is the reduced Planck constant.
Within the $\vec k\cdot\vec p$ approximation, the presence of an 
external electric field is handled by introducing the potential energy $U(\vec r)$ into Eq.~\eqref{dirac0}:
\begin{equation}\label{dirac}
\begin{aligned}
\displaystyle 
[ -i\gamma(\partial_x\sigma_x+\partial_y\sigma_y)+U(\vec{r})I ]\vec{F}^{\vec{K}} &= E\vec{F}^{\vec{K}} \\
\displaystyle 
[ -i\gamma(\partial_x\sigma_x-\partial_y\sigma_y)+U(\vec{r})I ]\vec{F}^{\vec{K'}} &= E\vec{F}^{\vec{K'}}\,,
\end{aligned}
\end{equation}
where $I$ is the $2\times 2$ identity matrix. 

Dirichlet boundary conditions for the wave function $\psi$ have to be imposed on the two dimer lines just 
outside the ribbon, at a distance $a/2$ from the ribbon edges, where passivation
approximately takes place. We  
choose the origin of the $y$ axis in such a 
way that these dimer lines are identified by the conditions $y=0$ and $y=\tilde W\equiv W+a$. 
The vanishing of the wave function on the passivation lines leads to the boundary 
conditions~\cite{brey2006,castroneto2009,kp2011}
\begin{equation}\label{boundary0}
\begin{aligned}
&\psi_{\beta} (x,y=0)=\psi_{\beta} (x,y=\tilde W)=0
\end{aligned}
\end{equation}
for both sublattices ($\beta=A,B$).

In a waveguide-like configuration in which the potential energy $U$ depends only on the transverse coordinate $y$, 
the longitudinal component of the momentum is constant; we will denote the longitudinal wave vector by $\kappa_x$.
The envelope functions that solve the Dirac equation \eqref{dirac} can be decomposed into a propagating wave
along $x$ and a confined component in the transverse direction:
\begin{equation}\label{Phi}
F_{\beta}^{\vec{\alpha}} (\vec{r})=
e^{i \kappa_x x}
\Phi_{\beta}^{\vec{\alpha}} (y)
\end{equation}
($\vec{\alpha}=\vec{K},\vec{K'}$; $\beta=A,B$). 
The functions $\Phi_\beta^{\vec\alpha}$ thus satisfy (\emph{cf.} Eq.~\eqref{dirac})
\begin{equation}\label{system}
\begin{aligned}
\displaystyle
\left[\sigma_x f(y)+\sigma_z\frac{d}{d y}\right]\vec{\varphi}^{\vec K}(y) &=-\kappa_x \vec{\varphi}^{\vec K}(y)\\
\displaystyle
\left[\sigma_x f(y)-\sigma_z\frac{d}{d y}\right]\vec{\varphi}^{\vec K'}(y) &=-\kappa_x \vec{\varphi}^{\vec K'}(y)\,,
\end{aligned}
\end{equation}
where we introduced the shorthands
\begin{equation}\label{varphik}
\vec\varphi^{\vec K}(y)=
\left[\begin{array}{c}
\Phi^{\vec K}_A(y)\\
\Phi^{\vec K}_B(y)
\end{array}\right]\, ,\quad
\vec\varphi^{\vec K'}(y)=
\left[\begin{array}{c}
i\,\Phi^{\vec K'}_A(y)\\
i\,\Phi^{\vec K'}_B(y)
\end{array}\right]\, ,
\end{equation}
and $f(y)=[U(y)-E]/\gamma$.

The boundary conditions \eqref{boundary0} become
\begin{equation} \label{boundary1}
\begin{aligned}
\displaystyle
\vec\varphi^{\vec K}(0) &=\vec\varphi^{\vec K'}(0)\\
\displaystyle
\vec\varphi^{\vec K}(\tilde W) &=e^{2 i K \tilde W} \vec\varphi^{\vec K'}(\tilde W)= e^{-i\eta\frac{2\pi}{3}}\vec\varphi^{\vec K'}(\tilde W)
\end{aligned}
\end{equation}
where
\begin{equation}
N_D+1\equiv \eta\!\!\!\mod 3
\end{equation}
and $\eta\in\{-1,0,1\}$. In particular, if $N_D+1=3M+\eta$ 
(with $M$ an integer), we have
\begin{equation} \label{n0}
\begin{aligned}
2 K \tilde W=2 \pi (2 M+\eta)-\eta\frac{2\pi}{3}\equiv 2 \pi n_0-\eta\frac{2\pi}{3}\, .
\end{aligned}
\end{equation}
The introduction of the discrete variable $\eta$ is convenient
since the product $K \tilde W$ can be very large in the case of wide ribbons,
where the envelope-function approximation is expected to be more reliable. 
In fact, $\eta$ is a geometrical property of the lattice structure that goes 
beyond the $\vec k\cdot \vec p$ approximation.

It is important to notice that although the envelope functions associated with different Dirac points decouple 
in the differential equations \eqref{system}, they are in fact mixed by the boundary conditions \eqref{boundary1}.
This coupling makes it nontrivial to define a symmetric discretization of \eqref{system}.
 
As previously noted, in the presence of a generic external electric field the differential eigenproblem \eqref{system}-\eqref{boundary1} 
cannot be solved analytically: it is necessary to rely on numerical methods in order to obtain approximate 
expressions for the transverse components $\Phi$ of the envelope functions and the
corresponding longitudinal wave vectors $\kappa_x$.

\section{Finite difference methods}\label{s:direct}

In this section we describe a few finite difference techniques
that could be adopted to numerically solve Eqs.~\eqref{system}-\eqref{boundary1}
in the direct space.

In a finite difference approach the unknowns are the values of the functions
$\Phi$ on a grid of $N_y$ points along the effective width $\tilde W$ of the 
ribbon; here we assume a uniform grid, by setting
$y_i=(i-1) \Delta_y$, with $\Delta_y=\tilde W/(N_y-1)$ and $i=1,\ldots, N_y$.
The derivatives are expressed as linear combinations of the values of the $\Phi$ on a finite number of grid
points, and the boundary conditions 
are constraints that reduce the number of unknowns.
As a consequence, the system of equations \eqref{system} is mapped to an algebraic
eigenvalue problem
$A \,\vec v=-\kappa_x \,\vec v$, where the elements of the vector
$\vec v$ are the values of the $\Phi$ at the grid points and the eigenvalues give the longitudinal wave 
vectors $\kappa_x$. 

In general, as $\Delta_y$ approaches zero, a subset of  eigenvectors of $A$, 
together with their respective eigenvalues, converge to the solutions of Eq.~\eqref{system}. 
The remaining eigenvectors and eigenvalues are discretization artifacts that have no meaningful continuum limit. 
We will refer to them as spurious solutions.

The adopted discretization scheme affects 
both the numerical efficiency and the appearance of spurious solutions. 
We now discuss the 
implications of the simplest discretization schemes:
\begin{enumerate}[(a)]
\item \label{e:asy}Naive asymmetric discretization 
\item \label{e:sym}Naive symmetric discretization 
\item \label{e:imp}Improved symmetric discretization.
\end{enumerate}

In scheme \eqref{e:asy} the first differential equation of \eqref{system} is evaluated at the
points $y_i$ with $i=1,\ldots,N_y-1$ and the second differential equation of \eqref{system} at $y_i$
with $i=2,\ldots,N_y$. Different representations for the derivatives are used: in the first equation
the two-point forward discretization formula $(d \Phi/d y)|_{y_i}\simeq [\Phi(y_{i+1})-\Phi(y_i)]/\Delta_y$ is used
and in the second one the two-point backward discretization formula $(d \Phi/d y)|_{y_i}\simeq 
[\Phi(y_i)-\Phi(y_{i-1})]/\Delta_y$.
The differential equations \eqref{system} are thus mapped to a
$4\,(N_y-1) \times 4\,(N_y-1)$ eigenvalue problem. 
We have also considered the alternative scheme of a symmetric discretization formula inside the 
ribbon and an asymmetric one at the edges.
In both cases a very slow convergence is observed. Moreover, a large 
number of spurious solutions are obtained, which persist also when higher order discretization schemes are used.

\begin{figure}
\begin{center}
\includegraphics[width=8cm]{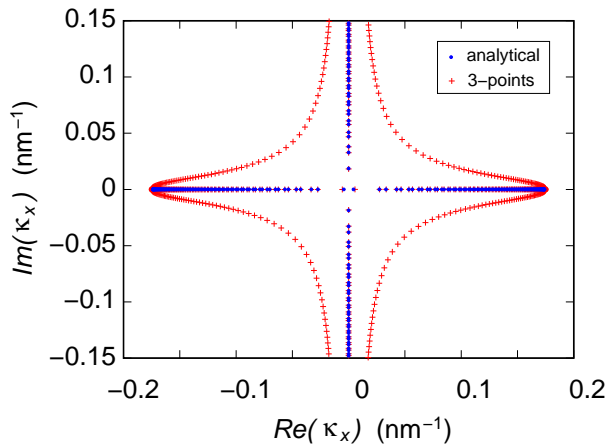}
\caption{(Color online) Eigenvalues $\kappa_x$ for a 
graphene nanoribbon with 8131 dimer lines ($\approx 1~\mu$m wide) and null 
potential energy, for a Fermi energy of 0.1~eV.
The analytical results are compared with those obtained with a 
standard three-point discretization scheme (symmetric inside the ribbon
and asymmetric at the edges), using a large number of discretization points
(5000) along the ribbon width.}
\label{figureig}
\end{center}
\end{figure}

As an example, in Fig.~\ref{figureig} we show the eigenvalues $\kappa_x$ 
obtained for $U(y)=0$ using a three-point 
discretization formula, symmetric inside the ribbon and asymmetric at 
the edges. 
This problem is analytically solvable, and the exact values 
of $\kappa_x$ turn out to be either real or purely imaginary
(see, e.g., Refs.~\cite{castroneto2009, kp2011}). 
The discretized problem instead also has a large number of complex 
solutions with nonzero real and imaginary parts (see Fig.~\ref{figureig}). 
In this simple case we can identify them as spurious solutions. However, for a 
generic potential energy function $U(y)$, 
complex solutions can be physical~\cite{PT}, so they cannot be rejected 
\emph{a priori}. If, instead of a three-point formula, we use, for 
example, a five-point formula for the spatial discretization, the nonspurious
eigenvalues converge onto the exact eigenvalues more quickly, as the 
discretization step is reduced, but the same number of spurious eigenvalues
are present.

In scheme \eqref{e:sym} the grid is modified to have a symmetric discretization in
every point of the grid. Defining $y_0=-\Delta_y$ and $y_{N_y+1}=\tilde W+\Delta_y$, the original
boundary conditions \eqref{boundary1} can be replaced with the relations
\begin{equation}\label{boundary2}
\begin{aligned}
\displaystyle
\vec{\varphi}^{\vec{K}}(y_0) &=\vec{\varphi}^{\vec{K'}}(y_1) \\
\displaystyle
\vec{\varphi}^{\vec{K'}}(y_0) &=\vec{\varphi}^{\vec{K}}(y_1) \\
\displaystyle
\vec{\varphi}^{\vec{K}}(y_{N_y+1}) &=e^{-i\eta\frac{2\pi}{3}}\vec{\varphi}^{\vec{K'}}(y_{N_y}) \\
\displaystyle
\vec{\varphi}^{\vec{K'}}(y_{N_y+1}) &=e^{i\eta\frac{2\pi}{3}}\vec{\varphi}^{\vec{K}}(y_{N_y})\,,
\end{aligned}
\end{equation}
which reduce to \eqref{boundary1} in the continuum limit.
Derivatives in \eqref{system} are evaluated by using the symmetric three-point discretization 
$(d\Phi/d y)|_{y_i}\simeq [\Phi(y_{i+1})-\Phi(y_{i-1})]/(2\,\Delta_y)$ in all the points
$y_i$ of the grid ($i=1,\ldots,N_y$) and the differential equations are thus mapped into a 
$(4\,N_y)\times (4\,N_y)$ eigenvalue problem. 

The eigenvalues of the discretized problem turn out to be always double degenerate.
In detail, each eigenspace is the span of two vectors, let us say  $\vec v_{(c)}$ and $\vec v_{(l)}$,
such that the components of $\vec v_{(l)}$
exhibit even-odd oscillations, while the others have oscillation frequency almost independent of $N_y$. 
The eigenvector $\vec v_{(l)}$ cannot have a continuum counterpart, hence the double degeneracy is in 
fact a lattice artefact. 

This is a clear manifestation of the so-called fermion doubling
problem: a ``naive'' direct space discretization of the Dirac equation results in the appearance of $2^d$ 
fermions (instead of one) in $d$ space dimensions (in our case $d=1$); this 
is an infrared effect, i.e. it does not disappear in the continuum limit $\Delta_y\to 0$. The fermion doubling is a very 
well known problem in the field of lattice quantum chromodynamics (see, e.g., Refs.~\cite{rothe, MM}) and is deeply 
connected with the chiral anomaly (see, e.g., Ref.~\cite{smilga}), i.e. with the impossibility of regularizing a 
theory with massless fermions in a local, chiral symmetric way.
In our simple case, it is caused by the symmetric three-point 
discretization formula for the derivative, which involves an incremental step of $2\,\Delta_y$, and hence 
decouples odd and even grid points.

Many methods have been developed to overcome the fermion doubling problem;
in scheme \eqref{e:imp} we employ the method proposed in Refs.~\cite{stacey1982, bender1983}, 
which has a quite simple implementation and was already applied in Ref.~\cite{tworzydlo2008}. 
The idea  
is to use a symmetric three-point discretization formula for the 
derivative, but with an incremental step equal to $\Delta_y$ instead of $2\,\Delta_y$. 
This can be done by evaluating the differential equations on
an auxiliary grid,  with nodes at the  
center coordinates $y_{i+(1/2)}$ of the cells of the original 
grid: $y_{i+(1/2)}=(y_i+y_{i+1})/2$, for $i=1,\ldots,N_y-1$.
The derivative is then approximated by
$(d\Phi/d y)|_{y_{i+(1/2)}}\simeq [\Phi(y_{i+1})-\Phi(y_i)]/\Delta_y$.
The potential energy $U(y)$ is known for every value of $y$ and can be directly evaluated at $y_{i+(1/2)}$, while the
value of the functions $\Phi$ at $y_{i+(1/2)}$ can be estimated by
the average of the values at $y_i$ and $y_{i+1}$:
$\Phi(y_{i+(1/2)}) \simeq [\Phi(y_{i})+\Phi(y_{i+1})]/2$.

The original differential equations \eqref{system} are thus mapped into the 
generalized algebraic eigenproblem 
$A \,\vec v=-\kappa_x \,B \,\vec v$,
with $A$ and $B$ that are $4\,(N_y-1) \times 4\,(N_y-1)$ matrices.
Since the matrix $B$ is invertible, this problem is in fact equivalent to the 
standard eigenproblem 
$(B^{-1} \,A) \,\vec v=-\kappa_x \,\vec v$. 
We notice, however, that while $A$ and $B$ are sparse 
matrices, $B^{-1}A$ is dense. As a consequence, optimized methods
to solve sparse eigenproblems (like the Arnoldi methods) cannot be directly 
applied to the standard form.
The sparsity of 
$A$ and $B$ can, however, still be exploited in 
the multiplication 
$\vec{y}=B^{-1}\, A\vec{x}$, which is the fundamental operation 
to be performed. 
This can be done by carrying out first
the multiplication by a sparse matrix $\vec{z}=A\vec{x}$ and then solving 
a sparse linear system
$B \vec{y}=\vec{z}$.
This discretization scheme solves the problems of schemes (\ref{e:sym}) and (\ref{e:imp}): there are neither 
spurious eigenvalues nor unphysical double degeneracies. 

In the next sections we will show that 
this is not the most 
efficient way to solve the system of differential equations \eqref{system}.

\section{Reformulation as a problem with periodic boundary conditions}\label{s:reformulation}

The numerical techniques in the real domain that we described in Sec.~\ref{s:direct} do not enable an efficient 
numerical analysis of the problem \eqref{system}-\eqref{boundary1}: in order to obtain 
high precision results, very large  matrices have to be diagonalized, and the size soon becomes prohibitive. 
In this section we reformulate the problem (\ref{system})-(\ref{boundary1}) on a different domain, but 
with periodic boundary conditions. In the next section we will show how to solve the resulting numerical problem in the 
reciprocal space.

We define the two-component function $\vec\varphi(y)$ by  
\begin{equation}\label{varphi}
\vec\varphi(y)=
\left\{
\begin{array}{l @{\quad} l}
\vec\varphi^{\vec K}(y) &
\hbox{ $y \in [0,\tilde W]$}\\
e^{-i\eta\frac{2\pi}{3}}\vec\varphi^{\vec K'}(2\tilde W-y) &
\hbox{ $y \in [\tilde W,2\tilde W]$}\, .
\end{array}
\right.
\end{equation}
From the second of the boundary conditions (\ref{boundary1}) we see that $\vec{\varphi}$ is continuous in its whole 
domain, while the first condition gives
\begin{equation}\label{perbc}
e^{-i\eta\frac{2\pi}{3}}\vec\varphi(0)=\vec\varphi(2\tilde{W})  \ . 
\end{equation}
Equation~\eqref{perbc} can be interpreted as the requirement 
of $2\tilde W$ periodicity 
for the function $\exp[i\eta \pi y/(3\tilde W)]\vec\varphi(y)$.

The differential equation satisfied by $\vec \varphi$ can be easily deduced 
from Eq.~(\ref{system}) and can 
be written in the compact form
\begin{equation}\label{system2}
\begin{aligned}
\displaystyle
\left[\frac{d}{d\,y} \sigma_z+h(y)\sigma_x\right]\vec\varphi(y) &=-\kappa_x \vec\varphi(y)\\
e^{i\eta\frac{2\pi}{3}}\vec\varphi(2\tilde W) &=\vec\varphi(0)\, ,
\end{aligned}
\end{equation}
where
\begin{equation}\label{h}
h(y)=f(\tilde{W}-|\tilde{W}-y|) \qquad  y \in [0,2\,\tilde W] \, .
\end{equation}

In this way we have halved the number of first-order differential equations by doubling the solution domain.
From Eq.~\eqref{system2} we see that 
$\kappa_x$ is an eigenvalue of the system [with corresponding
eigenfunction $\vec\varphi_{\kappa_x}(y)$], if and only if $-\kappa_x$, $\kappa_x^*$
and $-\kappa_x^*$ are eigenvalues, as well. 
The corresponding eigenfunctions can 
be expressed in terms of the eigenfunction of $\kappa_x$:
\begin{equation}\label{eq:Z2}
\begin{aligned}
\displaystyle
\vec\varphi_{-\kappa_x}(y) 
&\propto \sigma_y \,\vec\varphi_{\kappa_x}(y)\\
\displaystyle
\vec\varphi_{\kappa_x^*}(y) 
&\propto \sigma_x \left[\vec\varphi_{\kappa_x}(2\tilde W-y)\right]^*\\
\displaystyle
\vec\varphi_{-\kappa_x^*}(y) 
&\propto \sigma_z \left[\vec\varphi_{\kappa_x}(2\tilde W-y)\right]^*\, .
\end{aligned}
\end{equation}

It can be shown~\cite{PT} that there is no degeneracy when $\eta=\pm 1$, i.e. when $N_D+1$ is not 
divisible by three, while this is not generally true for $\eta=0$.
We also note that the solutions of the problems with $\eta=-1$ and $\eta=+1$ can be mapped into
each other by the relation
\begin{equation}
\vec\varphi_{k_x}^{\,(\eta=-1)}(y)=\sigma_x\vec \varphi_{k_x}^{\,(\eta=1)}(2\tilde W-y)\, .
\end{equation}
 
Solutions with non-real $\kappa_x^2$ (i.e. $\kappa_x$ with nonzero real and imaginary parts)
can be found in the presence of an external electric field. 
Their appearance is related to the existence of exceptional points, i.e. points in which 
the operator in Eq.~\eqref{system2} is not diagonalizable (notice that the operator is not self-adjoint). 
The existence of nonreal $\kappa_x^2$ values is a manifestation of the 
$\mathcal{PT}$ symmetry breaking in the system~\cite{PT}.

The methods described in the previous section [in particular, scheme \eqref{e:imp}] could also  be used to solve 
problem (\ref{system2}). However, we 
did not observe any significant efficiency gain with respect to 
the discretization of the original differential problem.

In the Appendix we show that Eq.~\eqref{system2} can 
be recast into the form of a complex second-order differential equation for a scalar unknown function. 
In that form the discretization in direct space is free from fermion doubling
effects. We did not
actively investigate its numerical solution, since the method 
that we are going to 
describe in the next section 
turns out to be much more efficient than the direct space ones~\cite{BF:2013}.

\section{Solution in the reciprocal space}
\label{s:fourier}

A key feature of the discretization in the direct space (discussed in Sec.~\ref{s:direct}) was the 
representation of the derivative.  

Let us consider a uniform grid with node spacing $a$ and the $n$-point discretization on it of the first 
derivative $\phi^{\prime}(p)$ of a generic function $\phi$, computed at point $p$. 
This discretization is constructed by Taylor expanding $\phi(p+i a)$ for different values of the integer $i$ 
and finding the linear combination of these expansions that is equal to $a\phi^{\prime}(p)$ up to 
$a^{n}\phi^{(n)}(p)$ corrections. If the function $\phi$ is smooth, the discretization error of the 
$n-$point derivative is thus $\mathcal{O}(a^{n-1})$ for every 
$n$. However, if the $\alpha$th order 
derivative of $\phi$ in $p$ is discontinuous, we cannot improve the 
precision of the discretization for $n>\alpha$ without introducing coefficients that depend on the specific function $\phi$ itself. 
As a consequence, the discretization error of a generic $n-$point discretization of the derivative scales as
\begin{equation}\label{eq:errordis}
|\phi^\prime (p)-\phi^{\prime}_{(N_y)}(p)|\gtrsim \mathcal O(N_y^{1-\min(\alpha,n)})\, ,
\end{equation}
where $\phi^{\prime}_{(N_y)}(p)$ is the $n$-point discretization of the first derivative on a grid
with $N_y$ points.

In the specific case~\eqref{system2}, it is enough that the first derivative of the potential is nonzero 
at the boundaries (i.e., the external electric field has a nonzero transverse component at the 
edges) for the second derivative of the eigenfunctions to be discontinuous at $y=0,\tilde W$ 
[see Eq.~\eqref{h}];  
if so, the accuracy of the approximation is 
independent of $n$ for $n\geq 2$,
and Eq.~\eqref{eq:errordis} represents a very severe limitation 
both for the precision and for the efficiency of the numerical solution.

The Fourier methods are better behaved in this respect. While the direct space 
methods involve a global distortion of the dispersion relation, 
in the Fourier case the derivative is exactly reproduced for
the frequencies lower than the cutoff.
The fermion doubling problem is 
absent, as one can argue by tracing back its origin to the periodicity of the wave function across the Brillouin 
zone induced by the space discretization (a simple topological argument is given in 
Ref.~\cite{smilga}, \S 13.1).

Since both $h(y)$ and $\exp[i\eta \pi y/(3\tilde W)]\vec\varphi(y)$ (separately) assume the same value at 0 and $2 \tilde W$,
they can be extended by periodicity with period $2 \tilde W$ without introducing discontinuities.
We define their Fourier coefficients $h_\ell\equiv h_{-\ell}$ and $\vec a_m$ by
\begin{equation}\label{expansion1}
\begin{aligned}
\displaystyle
h(y) &=\sum_{\ell=-\infty}^{\infty} h_\ell e^{i \pi \ell y /\tilde W} \, , \\ 
\displaystyle
\vec\varphi(y) &=
\sum_{m=-\infty}^{\infty}\vec a_m e^{i \pi(m-\eta/3) y /\tilde W} \, .
\end{aligned}
\end{equation}
We substitute these expressions in the differential equation of \eqref{system2} and then project
onto the exponential functions $e^{i \pi(n-\eta/3) y /\tilde W}$; for the 
generic index $n$ we obtain
\begin{equation}\label{eqfourier}
\sum_{m=-\infty}^{+\infty} 
\left[i \frac{\pi}{\tilde W}\left(n-\frac{\eta}{3}\right)\sigma_z \delta_{n,m}+
h_{n-m}\,\sigma_x \right]\vec a_{m}=-\kappa_x \vec a_n \, ,
\end{equation}
where $\delta_{n, m}$ is the Kronecker delta function. These equations are still exact and can be 
rewritten in the matrix form
\begin{equation}\label{eq:M}
M \vec{a}=-\kappa_x\vec{a}
\end{equation}
where $M$ is a structured infinite matrix whose $2\times 2$ block is given by
\begin{equation}\label{eq:M2x2}
M_{n,m}=P_n \delta_{n,m}+Q_{n,m}
\end{equation}
with
\begin{equation}
P_n=i \frac{\pi}{\tilde W}\left(n-\frac{\eta}{3}\right)\sigma_z \, , \quad 
Q_{n,m}=h_{n-m}\,\sigma_x \, .
\end{equation}

While the weight of the diagonal blocks $P_n$ increases with $|n|$,  
Parseval's theorem~\cite{korner1988} 
ensures that if $h(y)$ is square 
integrable the contribution of the blocks $Q_{n,m}$ vanishes for large values 
of $|n-m|$,
i.e., sufficiently far from the principal diagonal of the matrix $M$.
Actually, for the regular potentials for which the envelope
function approach gives reliable results, the hypothesis of square integrability
of $h(y)$ is a very weak assumption. 

If we consider a sufficiently large positive integer $D$ such that
\begin{equation}
\frac{\pi D}{\tilde W}\gg \max_j |h_j| \, ,
\end{equation}
the matrix $M_0$, with blocks 
$[M_0]_{n,m}=M_{n, m}$ and $|n|,|m|\leq D$, contains the main information about 
the slowly varying solutions of \eqref{eqfourier}: the Fourier coefficients that describe the low-frequency components of the spectrum
are well approximated by those of the truncated problem~\cite{BF:2013}; 
the high-frequency components 
are instead negligible for the slow varying solutions.
The finite dimensional problem
$(M_0 +\kappa_x I)\vec v_0=0$ is not affected by doubling, since it is just a 
truncation of the original problem \eqref{eqfourier}.

The eigenvalues of $M_0$ are accurate estimates of the longitudinal wave
vectors $\kappa_x$. Each eigenfunction $\vec\varphi(y)$ can be reconstructed,
using the corresponding eigenvector $\vec{a}_0$ of $M_0$, as  
\begin{equation}\label{antitransform}
\vec\varphi(y)\approx \vec\varphi_D(y)\equiv \sum_{\mu=-D}^{D}[\vec a_0]_\mu \,
e^{i \pi(\mu-\eta/3) y/\tilde W} \, .
\end{equation}
Using (\ref{varphi}) and (\ref{varphik}), the transverse components $\Phi$ of the envelope functions are given by ($\beta=A,B$)
\begin{equation}\label{Phi-an}
\begin{aligned}
\displaystyle
\Phi^{\vec K}_\beta (y) &\approx
\sum_{\mu=-D}^{D} [a_0^\beta]_\mu \,e^{i \pi (\mu-\eta/3) y /\tilde W}\, ,\\
\displaystyle
\Phi^{\vec K'}_\beta (y) &\approx-i
\sum_{\mu=-D}^{D} [a_0^\beta]_\mu \,e^{-i \pi (\mu-\eta/3) y /\tilde W}\, .
\end{aligned}
\end{equation}
Finally, from \eqref{psia} and \eqref{Phi} it follows that
\begin{equation} \label{combisin}
\psi_\beta (x,y)=2\,i \sum_{\mu=-D}^D 
\left\{ [a_0^\beta]_\mu \,\sin[(\mu-n_0) \pi y / \tilde W ]\right\}
e^{i \kappa_x x} \, ,
\end{equation}
where  $n_0$ has been defined in \eqref{n0}.

We explicitly note that, from the numerical point of view, all these computations 
strongly benefit from the use of optimized fast Fourier transform routines 
for the calculation of the Fourier series.

\section{Numerical efficiency: comparison among methods}\label{s:comparison}

In this section we compare the numerical efficiency of the methods 
introduced so far, performing an analysis of the convergence rate 
for several test cases.

We are going to compare three main strategies:
\begin{enumerate}
\item[$(S)$] method \eqref{e:imp} of Sec.~\ref{s:direct}
\item[$(S_p)$] method \eqref{e:imp} of Sec.~\ref{s:direct} applied to the periodic problem \eqref{system2}
\item[$(F)$] the Fourier method described in Sec.~\ref{s:fourier}.
\end{enumerate}

\begin{figure}
\begin{center}
\includegraphics[width=8cm]{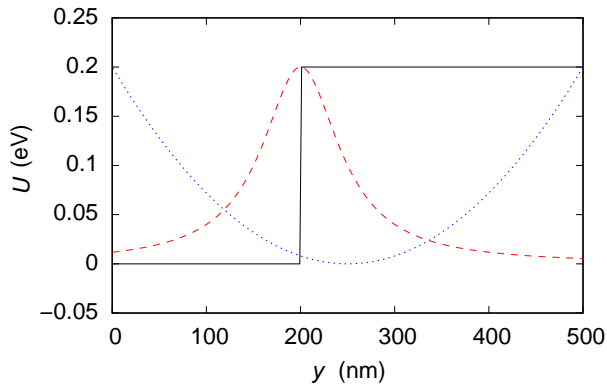}
\caption{(Color online) Plot of the potentials used to test the 
numerical efficiency of the 
methods: a step potential (solid curve), a Lorentzian potential 
(dashed curve), and a parabolic potential
(dotted curve).}
\label{pot}
\end{center}
\end{figure}

We consider a nanoribbon composed  of \mbox{$N_D=4065$} dimer lines (corresponding to $\eta=1$ and to an effective 
width \mbox{$\tilde{W}\approx 500\,\mathrm{nm}$}), with the following potentials (shown in Fig.~\ref{pot}):
\begin{itemize}
\item[(1)] Step potential 
\begin{equation}
U(y)=\left\{\begin{array}{ll}
0\,\mathrm{eV} \qquad & y\le 200\,\mathrm{nm} \\
0.2\,\mathrm{eV} & y>200\,\mathrm{nm}
\end{array}\right.
\end{equation}
\item[(2)] Lorentzian potential 
\begin{equation}
U(y)=A\frac{\Gamma/2}{(y-y_0)^2+(\Gamma/2)^2}
\end{equation}
with $y_0=200\,\mathrm{nm}$, $\Gamma=100\,\mathrm{nm}$ and $A=10\,\mathrm{eV}\,\mathrm{nm}$
\item[(3)] Parabolic potential 
\begin{equation}
U(y)=\bar{A}(y-\bar{y})^2
\end{equation}
with $\bar{y}=250\,\mathrm{nm}$ and $\bar{A}=0.2\,\mathrm{eV}/(250 \,\mathrm{nm})^2$\, .
\end{itemize}
We set the electron injection energy to $E=0.1\,\mathrm{eV}$ and study the scaling of the eigenvalue precision as a 
function of the execution time on an Intel Xeon CPU E5420 2.50GHz processor. Diagonalization is performed 
by means of standard LAPACK routines. In case $(F)$, the coefficients $h_n$ are computed  
on an extremely fine grid, independent of the dimension $D$ of the truncated problem, without introducing any 
sizable overhead. 

Figures refer to the maximum real eigenvalues. We did not find 
significant differences in the behavior of the other eigenvalues. 
However, eigenvalues 
associated with larger $\kappa_x$ values converge faster to the corresponding 
eigenvalues of the original problem~\eqref{system2}. 
This can be interpreted as a consequence of the fact that large $\kappa_x$ values correspond to a small kinetic energy 
in the transverse direction 
(the total energy is constant), i.e., to transverse modes with large wavelength, which are less sensitive to the 
discretization or the frequency cutoff.

\begin{figure}
\begin{center}
\includegraphics[width=8cm]{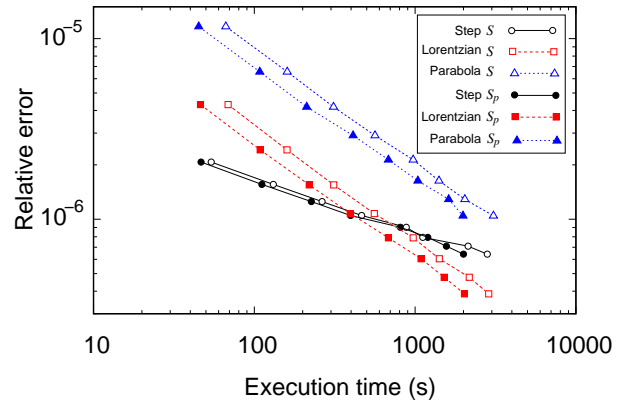}
\caption{(Color online) Relative error on the largest real 
eigenvalue as a function of execution time, when using the spatial methods 
$S$ and $S_p$ (see text for the definition of abbreviations). Different
execution times correspond to different discretization steps.}
\label{ssp}
\end{center}
\end{figure}

\begin{figure}
\begin{center}
\includegraphics[width=8cm]{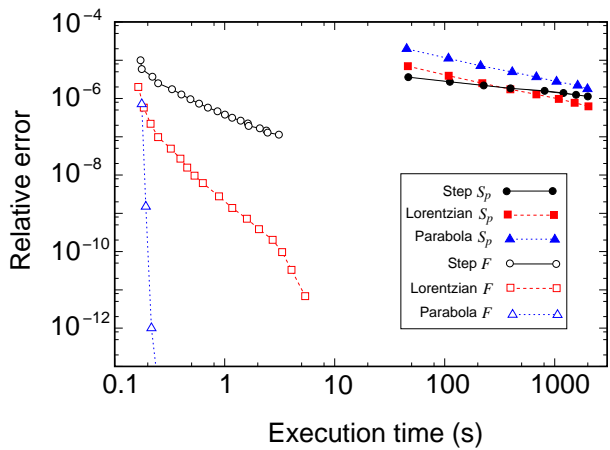}
\caption{(Color online) Relative error on the largest real eigenvalue as a 
function of the execution time. Comparison between the Fourier method and the 
spatial $S_p$ one (see text for the definition of abbreviations). In the case 
of the parabolic potential the precision very quickly reaches the machine 
precision. Different execution times correspond to different values of the 
discretization step for the $S_p$ method and to different numbers of 
components for the Fourier method.}
\label{fsp}
\end{center}
\end{figure}

In Figs.~\ref{ssp} and \ref{fsp} we report the relative error on the largest 
real eigenvalue for the two spatial approaches $S$ and $S_p$ 
(in Fig.~\ref{ssp}), and the methods $S_p$ and $F$ (in Fig.~\ref{fsp}) as
a function of the execution time. The data points for different 
values of the execution time have been obtained by varying the discretization 
step in the case of the spatial methods, and varying the number of considered 
Fourier components in the case of the Fourier methods (the smaller the 
discretization step or the greater the number of Fourier components, the
larger the execution time).
Method $(S_p)$ is slightly more efficient than method $(S)$, probably due to the better block structure of the 
discretization matrix. However, in all the cases we have studied the Fourier methods largely outperform the direct space 
ones, often by several orders of magnitude.

Figure \ref{fsp} also shows that the convergence of the 
Fourier method
is strongly dependent on the shape and the analytic properties of the 
potential, which   
influence the number of Fourier coefficients needed to properly expand
the eigenfunctions (and, in turn, the size of the matrices to be 
diagonalized).

The better performance of Fourier methods with respect to direct space ones has been recently noticed also 
in Ref.~\cite{wolff2012}, where a Schr\"odinger equation with position dependent mass is considered. 
The authors used the following ``Fourier-inspired'' discretization for the derivative ($\mathcal{F}$ is the discrete 
Fourier transform and $k$ is the reciprocal space variable)
\begin{equation}
\frac{d}{d x}\longrightarrow \mathcal{F}^{-1}\,k\,\mathcal{F}
\end{equation}
and reported a convergence rate exponentially fast in the number of points of the discrete Fourier transform.

If we compare the numerical errors of the methods as a function of the size of the matrices involved in the analysis, we conclude that 
the errors deriving from the finite-difference discretization of the 
derivatives (and from the resulting distortion of the
dispersion relation) turn out to be much larger than those related 
to the cutoff of the high-frequency Fourier components in the reciprocal 
space approach.
This shows that, for the type of potentials we are interested in, the Fourier method is drastically 
more efficient than the others.

\section{Solution of the transport problem}\label{s:transport}

In the previous sections we have described numerical methods to compute the eigenvalues and the eigenfunctions 
of the Dirac equation in a longitudinally invariant ribbon.
The total wave functions on the two sublattices $\psi_\beta (\vec r)$ ($\beta=A,B$) are a linear 
combination of modes $\psi_{\beta i} (\vec r)$ of the form:
\begin{equation}\label{chi}
\begin{aligned}
\psi_{\beta i} (\vec r) &=
\left[e^{-iKy} \Phi^{\vec K}_{\beta i} (y)-ie^{iKy} 
\Phi^{\vec K'}_{\beta i} (y)\right]
e^{i \kappa_{xi} x} \\
&\equiv\chi_{\beta i} (y)\,e^{i \kappa_{xi} x}\ .
\end{aligned}
\end{equation}

In the case of a general potential $U(\vec{r}\,)$ 
we divide the ribbon into a series of transverse slices, in such a way that within each slice 
the potential is approximately independent of $x$.
For each slice we can then apply the previously discussed methods to estimate the modes $\psi_{\beta i}$ and their 
longitudinal momenta $\kappa_x$. 
At the interfaces between adjacent slices we have to enforce the continuity of the total wave 
function (the $\vec{k}\cdot\vec{p}$ approximation is reliable only if the potential varies slowly
on the lattice scale, thus no $\delta$-type potentials are allowed).
We remark here a difference with respect to the standard 
Schr\"odinger case: the Dirac equation is a first-order differential 
equation, thus we do not have to impose the continuity of  the normal 
derivative of the wave function. 

Since the atomic orbitals in Eq.~\eqref{total} are strongly localized, 
enforcing the continuity of the total wave function 
amounts to imposing the continuity of the wave functions on both inequivalent 
sublattices separately.
Moreover, since the functions $\psi_\beta (\vec r\,)$ 
have Fourier components localized around the two inequivalent Dirac
points (which are significantly separated from each other), the continuity of 
the functions $\psi_\beta (\vec r\,)$ implies also the continuity of the 
envelope functions $F$.

Integrating the probability current density 
in the $x$ direction~\cite{kp2011} 
\begin{equation}\label{jx}
J_x(y) =v_F\,[
\vec{F}_{\vec{K}}^{\dag}(y)\sigma_x\vec{F}_{\vec{K}}(y)+
\vec{F}_{\vec{K}'}^{\dag}(y)\sigma_x\vec{F}_{\vec{K}'}(y)]
\end{equation}
over the transverse section, and using Eq.~(\ref{Phi-an}),
we can express the longitudinal probability current as follows:
\begin{equation}
\begin{aligned}
I_x &=\int_0^{\tilde W} J_x(y)d y
=4v_F \tilde W\,\Re\left[\sum_{n=-D}^D {(a_n^A)}^* a_n^B\right] \\
&=v_F\,\int_0^{2\tilde{W}} \vec{\varphi}(y)^{\dag}\sigma_x\vec{\varphi}(y)d y \, .
\end{aligned}
\end{equation}

From Eq.~\eqref{eq:Z2} we deduce that $(I_x)_{-\kappa_x}=-(I_x)_{\kappa_x}$, 
$(I_x)_{\kappa_x^*}=(I_x)_{\kappa_x}$ and $(I_x)_{-\kappa_x^*}=-(I_x)_{\kappa_x}$.
In particular, if $\kappa_x$ is purely imaginary, we have 
$I_x=0$, i.e., modes with purely imaginary eigenvalues do not carry
current. This is in general not true for eigenvalues that have at the same time
a nonzero real and imaginary part.
We classify the modes as right-moving or left-moving, depending on whether they have a 
positive or negative longitudinal probability current $I_x$. We also extend the definition of 
right-moving (left-moving) to the modes with $I_x=0$ and $\Im(\kappa_x)>0$ ($\Im(\kappa_x)<0$).

In detail, in our simulation code we order the modes
on the basis of the value of the corresponding $\kappa_x$. We first
consider the real $\kappa_x$ (arranged in order of decreasing modulus),
then the complex ones, and finally the purely imaginary ones (sorted in
order of increasing modulus). Since this ordering reflects the expected weight
of the different modes in a transport simulation, in our computations we  
consider only the first 
$n_{mod}$ right-moving modes and the first $n_{mod}$ left-moving modes of each slice.
Clearly $n_{mod}$ has to be large enough for the final physical result to be insensitive
to its specific value. Moreover, we select the modes in such a way as to preserve the 
$Z_2\times Z_2$ symmetry, which means to pick at 
the same time the modes with eigenvalues $\kappa_x$, $-\kappa_x$,
$\kappa_x^\ast$, and $-\kappa_x^\ast$. 

Let us now sketch the basic steps to compute the scattering matrix 
for a single discontinuity of the potential at the interface between adjacent 
slices.
We denote by $l/r$ the modes on the left/right of the discontinuity and 
by $+/-$ the right/left-moving modes. 
We use the index $i$ to denote the mode impinging on the discontinuity, e.g., from the left.
The wave function $\psi_\beta (\vec r)$ on the left side can be written as
\begin{equation}\label{eq:left}
\chi^{l+}_{\beta i} (y)\,e^{i \kappa^{l+}_{xi} (x_{dis}-x_{in})}
+\sum_n r_{ni} \chi^{l-}_{\beta n} (y)\,e^{i \kappa^{l-}_{xn} (x_{dis}-x_{in})} \ ,
\end{equation}
while on the right side it can  be expressed in the form
\begin{equation}\label{eq:right}
\sum_n t_{ni} \chi^{r+}_{\beta n} (y)\,e^{i \kappa^{r+}_{xn} (x_{dis}-x_{out})} \ . 
\end{equation}
Here $x_{in}$ and $x_{out}$ are the longitudinal positions of the
boundaries of the considered scattering region, $x_{dis}$ is the position of the discontinuity, while
$r_{ni}$ and $t_{ni}$ are the reflection and transmission coefficients.
By continuity, functions \eqref{eq:left} and \eqref{eq:right} must be equal.
An analogous relation can be established for a mode injected from
the right.
These continuity relations have to be enforced for both sublattices and for
all the $2\,n_{mod}$ modes impinging from the left and from the right.

In order to evaluate all the $4\,n_{mod}^2$ reflection and transmission coefficients,
we can project the $4\,n_{mod}$ continuity constraints onto a set of functions chosen
in such a way as to obtain the correct number of independent equations.
From Eq.~\eqref{combisin} we have
\begin{equation}
\chi_{\beta i} (y)=2\,i \sum_{n=-D}^D 
\left\{ a_n^{\beta i}\,\sin[(n-n_0) \pi y / \tilde W ]\right\} \, ,
\end{equation}
hence it is natural to project each continuity relation
on the set of $n_{mod}$ functions
\begin{equation}\label{sinfunc}
S_j (y)=\sin\left((j-n_0) \pi y / \tilde W \right)\ ,
\end{equation}
for $j=-(n_{mod}-1)/2 , \ldots, (n_{mod}-1)/2$
(for the sake of simplicity $n_{mod}$ is assumed odd).
Since we consider values of $n_{mod}$ such
that $(n_{mod}-1)/2 < n_0$, these functions are linearly independent. 
It is simple to show that the matrix elements are 
\begin{equation}
\langle S_j (y) | \chi_{\beta i} (y) \rangle =
\int_0^{\tilde W} S_j^* (y) \chi_{\beta i} (y) d y=
i \tilde W a_j^{\beta i} \, ;
\label{samew}
\end{equation}
thus all the computations can be performed in the reciprocal space, 
avoiding the evaluation of the sums in Eq.~\eqref{Phi-an}.

Once the scattering matrices corresponding to the various interfaces have been computed, 
they can be composed according to the standard procedure (see, e.g., Ref.~\cite{datta}) 
to obtain the total scattering matrix $S$ of the ribbon:
\begin{equation}
S=\left(\begin{array}{cc}
r & \tilde{t} \\
t & \tilde{r} 
\end{array}\right)
\end{equation}
Here $r$ and $t$ are the reflection and transmission matrices for the modes
impinging from the left, $\tilde{r}$ and $\tilde{t}$ the corresponding 
matrices for the modes
impinging from the right.

For practical purposes it is convenient to introduce the current form $S'$ of the scattering
matrix:
\begin{equation}
S'=\left(\begin{array}{cc}
r' & \tilde{t}' \\
t' & \tilde{r}' 
\end{array}\right) \, ,
\end{equation}
relating the ``current amplitudes'' instead of the ``wave amplitudes'' of
the modes~\cite{datta}.
This matrix involves only the modes with $I_x\neq 0$ and its elements are
given by $s'_{nm}=s_{nm}\sqrt{|I_{xn}|/|I_{xm}|}$, with $s=r,t,\tilde r,\tilde t$.
As a result of current conservation, it can be shown that $S'$ is unitary
(see, e.g., Ref.~\cite{datta}), which is a useful check to be performed at the end of the computations.
In all our simulations we checked that numerical violations of the unitarity relation are less
than $10^{-13}$.

\begin{figure}[t]
\begin{center}
\includegraphics[width=8cm]{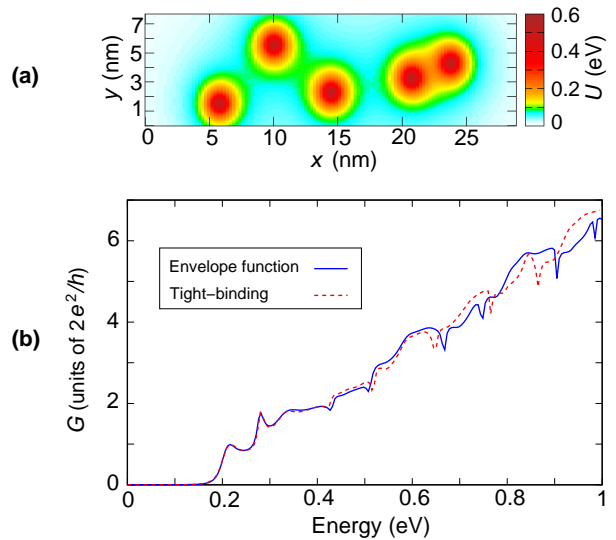}
\caption{(Color online) (a) Map of the potential in the nanoribbon,
given by a superposition of Lorentzian functions.
(b) Normalized conductance as a function of the injection energy,
obtained within the envelope function and the 
nearest-neighbor semiempirical tight-binding approximations.}
\label{fig5}
\end{center}
\end{figure}

\begin{figure}[t]
\begin{center}
\includegraphics[width=8cm]{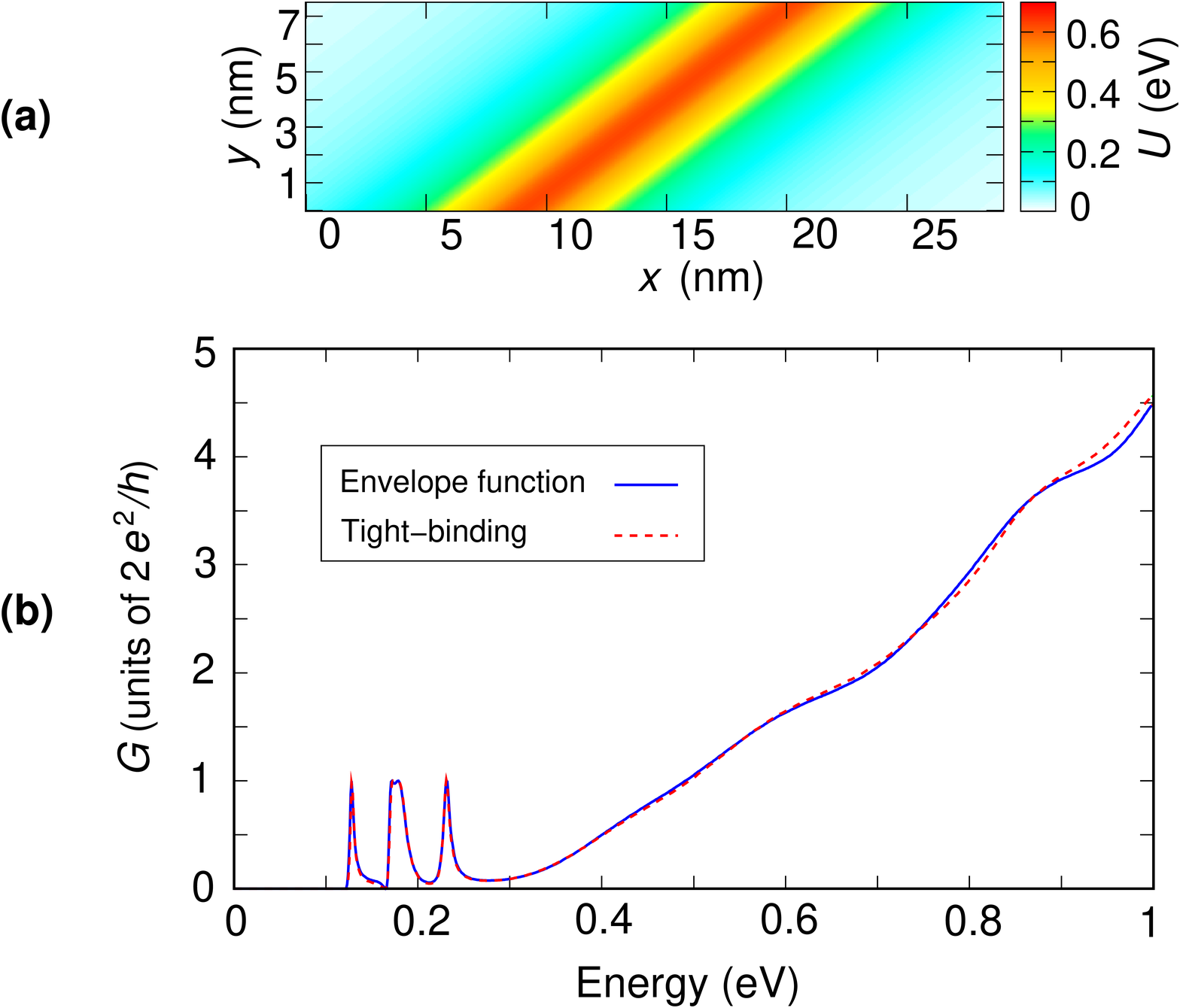}
\caption{(Color online) (a) Map of the potential in the nanoribbon,
represented by a tilted barrier with a Lorentzian profile.
\hfill\break
(b) Normalized conductance as a function of the injection energy,
obtained within the envelope function and the 
nearest-neighbor semiempirical tight-binding approximations.}
\label{fig6}
\end{center}
\end{figure}
 
From $S'$ we can compute the conductance of the ribbon by means of the
Landauer-B\"uttiker formula
\begin{equation}
G=\frac{2\,e^2}{h}\sum_{n,m} |t'_{nm}|^2\ ,
\label{landauer}
\end{equation}
where the sum runs over the modes with $I_x \ne 0$ in the first and last
transverse regions of the ribbon.

In studies of unconfined graphene or of ribbons with large aspect ratio, it is
usual to assume periodic instead of Dirichlet boundary conditions;
the two Dirac points are then completely decoupled, and it is customary to
solve the Dirac equation for just one valley and use a factor of four instead 
of two in Eq.~\eqref{landauer}. The physical Dirichlet boundary 
conditions introduce instead a coupling between the two inequivalent Dirac points,
requiring the use of the more general formulation~\eqref{landauer}.

For validation purposes, we have performed a transmission calculation for a structure that is small enough
to allow also a treatment with a standard tight-binding code
(in particular we have used NanoTCAD ViDES)~\cite{vides1,vides2}.
We considered an armchair nanoribbon with $60$ dimer lines 
($\approx 7.5$~nm wide), in the presence of two different 
realistic potential profiles.

In the first test case the electrostatic potential is a superposition of 
Lorentzian functions,
and can schematically represent the effect on the ribbon of charged impurities 
located in the substrate on which graphene lies.
In particular, we consider five Lorentzian functions, with a peak amplitude
of $0.5$~eV and a half-width at half-maximum equal to $0.64$~nm
(see Fig.~\ref{fig5}(a)).
In Fig.~\ref{fig5}(b) we show the computed behavior of the 
conductance $G$ (in units of $2 e^2/h$)
as a function of the injection energy, together with the corresponding 
results obtained with ViDES.
We observe a very good agreement between  
the two different approaches in the low injection energy regime $E_{in}\lesssim 0.5$~eV, i.e., the one in which 
the envelope function method can be safely applied. For larger energies,
the simple Dirac equation, which represents only a 
first-order $\vec k\cdot \vec p$ approximation,
does not appropriately describe the physics of graphene any more, and thus 
discrepancies between
the two results appear, even though the qualitative behavior 
of the conductance is well
reproduced for all the explored injection energies.

The other potential profile we have considered is
a barrier, tilted with respect to the ribbon edges and with a Lorentzian 
profile, with a peak amplitude of $0.625$~eV and a half-width at 
half-maximum equal to $2$~nm [see Fig.~\ref{fig6}(a)]. The ribbon has a 
width of 60 dimer lines, as in the previous example. This
potential can be the representation of the electrostatic effect at 
the graphene level of a biased gate or of a line of charge present at a 
certain distance from it. In Fig.~\ref{fig6}(b) we report
the behavior of $G$ (in units of $2 e^2/h$) as a function of the electron 
energy, 
obtained with our envelope-function-based calculation and with the 
tight-binding code. Also in this case we notice a good agreement 
between the two approaches, especially in the low-energies regime, 
in which the $\vec k\cdot \vec p$ approximation is expected 
to be more accurate.

Indeed, our numerical analysis is based on the use, in each section with 
longitudinally constant potential, of the continuum Dirac 
equation \eqref{dirac}. This approximation is valid if the following 
conditions are satisfied:
\begin{enumerate}
\item[(1)] The dispersion relation is approximately linear
\item[(2)] The wave function is slowly varying on the scale of the lattice spacing
\item[(3)] The potential is slowly varying on the scale of the lattice spacing.
\end{enumerate}

In the hypothesis of a slowly varying potential~\cite{ashcroft}, the energy of an
electron with wave vector $\vec{\kappa}$ in position $\vec{r}$ can be written 
as $E \simeq T(\vec{\kappa})+U(\vec{r}\,)$, where $T$ is the kinetic energy.
Observing the explicit form of $T(\vec{\kappa})$ (i.e., the actual
dispersion relation in the absence of potential 
energy, a relationship which can be derived, for
example, with a tight-binding formulation~\cite{castroneto2009, saito}), we see
that the first condition, i.e., the linear approximation for $T(\vec{\kappa})$
($T(\vec{\kappa})\simeq \pm\hbar v_F|\vec \kappa|$), is
valid for $|T(\vec{\kappa})|\lesssim 1~\mathrm{eV}$, which corresponds to
\begin{equation}\label{cond1}
|E-U(\vec{r}\,)|\lesssim 1~\mathrm{eV}\,.
\end{equation}

The second condition can be expressed in the form 
$\lambda=2\pi/|\vec \kappa|\gg a$ 
(where $\lambda$ is the electron wavelength). Exploiting the relation
(valid under the previous approximations)
$E\simeq \pm\hbar v_F|\vec \kappa|+U(\vec r)$,
this inequality translates into the condition 
\begin{equation}\label{cond2}
|E-U(\vec r)| \ll (2\pi\hbar v_F)/a\approx 15~\mathrm{eV}\, ,
\end{equation}
which is clearly weaker than Eq.~\eqref{cond1}.

Instead, the requirement on the smoothness of the potential (the third
condition) introduces a limitation on the derivative of $U$ along $y$
(the only spatial variable along which $U$ varies within each section). If 
we require the variation of the potential energy over the lattice constant 
$a$ to be negligible with respect to $\gamma_0$ [which represents an 
order of magnitude of the energies involved, since, in a first
approximation~\cite{saito}, $T(\vec{\kappa})$ has values between
$-3\gamma_0$ and $3\gamma_0$], this further constraint can be expressed as
\begin{equation}\label{cond3}
|\partial U/\partial y|\ll \gamma_0/a\approx 11~\mathrm{eV}/\mathrm{nm}\ .
\end{equation}
In the previous numerical examples, Eq.~\eqref{cond3} was always well 
satisfied; thus the only 
limit to the application of the Dirac equation was the condition 
\eqref{cond1} on the injection 
energy. Indeed, as previously noted, for $E\gtrsim 0.5$~eV the 
continuum and the tight-binding results
are not in as good an agreement as in the low-energy region.

\section{Conclusions}\label{s:conclusions}

We have presented a numerically efficient approach, including physical boundary conditions, for the evaluation of transport
properties of graphene devices for which the 
application of atomistic techniques is computationally prohibitive.
We have focused on ribbons with armchair edges, 
which we modeled within a continuum, envelope function approximation. 

For the computation of the transmission we have adopted a recursive scattering 
matrix approach, which requires the solution of a collection of Dirac equations 
in the presence of longitudinally constant potentials. 
We have shown that a reciprocal space approach is largely 
preferable with respect to
the more commonly adopted finite difference methods, since it can reduce
the computational cost of the procedure by orders of magnitude.

We have compared our results for structures small enough to allow an atomistic 
simulation with those obtained by means of tight-binding techniques, finding good agreement 
within the range of validity of the $\vec k\cdot\vec p$ approximation.

\section*{Acknowledgments}

It is a pleasure to thank Paolo Pintus for useful discussions.

\newpage

\appendix*
\section{ALTERNATIVE FORMULATION}
\label{appendix}

We  present here a reformulation of Eq.~(\ref{system2}) as 
a second order differential equation for a scalar function. Let $\xi(y)$ be defined by
\begin{equation}\label{equation1d-b}
\begin{aligned}
\displaystyle
\left[-\frac{d^2}{d\,y^2}+ 2i\, h(y) \frac{d}{d\,y}\right]\xi(y) &=-\kappa_x^2 \xi(y)\\
\displaystyle
\xi(2 \tilde W) &=e^{2 i K_0 \tilde W} \xi(0)\\
\displaystyle
\xi'(2 \tilde W) &=e^{2 i K_0 \tilde W} \xi'(0)
\end{aligned}
\end{equation}
where $\xi'(y)$ is a shorthand for $d \xi/d y$ and $K_0$ is defined as
\begin{equation}
K_0=K+\frac{1}{\tilde W}\int_0^{\tilde W} h(\alpha)d\alpha \, .
\end{equation}
One can easily verify that system (\ref{system2})  is solved by
\begin{equation}\label{equation1d-a}
\vec\varphi(y)=e^{-i \int_0^y h(\alpha) d \alpha}\left[\kappa_x \xi(y)\binom{1}{i}-\xi'(y)\binom{1}{-i}\right] \, .
\end{equation}


\begin{thebibliography}{99}

\bibitem{geim2004}
K.~S.~Novoselov, A.~K.~Geim, S.~V.~Morozov, D.~Jiang, Y.~Zhang, S.~V.~Dubonos, I.~V.~Grigorieva, and A.~A.~Firsov,
Science {\bf 306}, 666 (2004).

\bibitem{katsnelson2007}
M.~I.~Katsnelson and K.~S.~Novoselov,
Solid State Commun. {\bf 143}, 3 (2007).

\bibitem{geim2009}
A.~K.~Geim, 
Science {\bf 324}, 1530 (2009).

\bibitem{iannaccone2009}
G.~Iannaccone, G.~Fiori, M.~Macucci, P.~Michetti, M.~Cheli, A.~Betti, and P.~Marconcini,
in IEEE Int. Electron Device Meeting (IEDM), IEEE Conference Proceedings, 245 (2009),
DOI: 10.1109/IEDM.2009.5424376.

\bibitem{mccann2006}
E.~McCann and V.~I.~Fal'ko,
Phys. Rev. Lett. {\bf 96}, 086805 (2006).

\bibitem{ohta2006}
T.~Ohta, A.~Bostwick, T.~Seyller, K.~Horn, and E.~Rotenberg,
Science {\bf 313}, 951 (2006).

\bibitem{castro2007}
E.~V.~Castro, K.~S.~Novoselov, S.~V.~Morozov, N.~M.~R.~Peres, J.~M.~B.~Lopes dos Santos, J.~Nilsson,
F.~Guinea, A.~K.~Geim, and A.~H.~Castro~Neto,
Phys. Rev. Lett. {\bf 99}, 216802 (2007).

\bibitem{zhang2009}
Y.~Zhang, T.-T.~Tang, C.~Girit, Z.~Hao, M.~C.~Martin, A.~Zettl, M.~F.~Crommie, Y.~R.~Shen, and F.~Wang, 
Nature (London) {\bf 459}, 820 (2009).

\bibitem{ouyang2008}
Y.~Ouyang, Y.~Yoon, and J.~Guo, 
in IEEE Int. Electron Device Meeting (IEDM), Tech. Dig., 4796739 (2008), DOI: 10.1109/IEDM.2008.4796739.

\bibitem{echtermeyer2009}
T.~J.~Echtermeyer, M.~C.~Lemme, M.~Baus, B.~N.~Szafranek, A.~K.~Geim, and H.~Kurz, 
IEEE Electron Device Lett. {\bf 29}, 952 (2009).

\bibitem{elias2009}
D.~C.~Elias, R.~R.~Nair, T.~M.~G.~Mohiuddin, S.~V.~Morozov, P.~Blake, M.~P.~Halsall,
A.~C.~Ferrari, D.~W.~Boukhvalov, M.~I.~Katsnelson, A.~K.~Geim, and K.~S.~Novoselov,
Science {\bf 323}, 610 (2009).

\bibitem{boukhvalov2009}
D.~W.~Boukhvalov and M.~I.~Katsnelson,
J. Phys.: Condens. Matter {\bf 21}, 344205 (2009).

\bibitem{acs2012}
P.~Marconcini, A.~Cresti, F.~Triozon, G.~Fiori, B.~Biel, Y.-M.~Niquet, M.~Macucci, and S.~Roche,
ACS Nano {\bf 6}, 7942 (2012).

\bibitem{iwceroche}
P.~Marconcini, A.~Cresti, F.~Triozon, G.~Fiori, B.~Biel, Y.-M.~Niquet, M.~Macucci, and S.~Roche,
in Proceedings of IWCE 2012, IEEE Conference Proceedings, 6242844 (2012),
DOI: 10.1109/IWCE.2012.6242844.

\bibitem{tang2012}
Y.-B.~Tang, L.-C.~Yin, Y.~Yang, X.-H.~Bo, Y.-L.~Cao, H.-E.~Wang, W.-J.~Zhang, I.~Bello, S.-T.~Lee, H.-M.~Cheng, and C.-S.~Lee,
ACS Nano {\bf 6}, 1970 (2012).

\bibitem{wei2009}
D.~Wei, Y.~Liu, Y.~Wang, H.~Zhang, L.~Huang, and G.~Yu,
Nano Lett. {\bf 9}, 1752 (2009).

\bibitem{novoselov2012}
K.~S.~Novoselov, V.~I.~Fal'ko, L.~Colombo, P.~R.~Gellert, M.~G.~Schwab, and K.~Kim,
Nature (London) {\bf 490}, 192 (2012).

\bibitem{schwierz2010}
F.~Schwierz,
Nature Nanotech. {\bf 5}, 487 (2010).

\bibitem{mina2013}
A.~N.~Mina and A.~H.~Phillips,
J. Appl. Sci. Res. {\bf 9}, 1854 (2013).

\bibitem{hill2011}
E.~W.~Hill, A.~Vijayaragahvan, and K.~Novoselov,
IEEE Sensors J. {\bf 11}, 3161 (2011).

\bibitem{bonaccorso2010}
F.~Bonaccorso, Z.~Sun, T.~Hasan, and A.~C.~Ferrari,
Nature Photon. {\bf 4}, 611 (2010).

\bibitem{xia2009}
F.~Xia, T.~Mueller, Y.-m.~Lin, A.~Valdes-Garcia, and Phaedon Avouris,
Nature Nanotech. {\bf 4}, 839 (2009).

\bibitem{cooper2012}
D.~R.~Cooper, B.~D'Anjou, N.~Ghattamaneni, B.~Harack, M.~Hilke, A.~Horth,
N.~Majlis, M.~Massicotte, L.~Vandsburger, E.~Whiteway, and V.~Yu,
ISRN Condensed Matter Physics {\bf 2012}, 501686 (2012),
DOI: 10.5402/2012/501686.

\bibitem{peres2010}
N.~M.~R.~Peres, 
Rev. Mod. Phys. {\bf 82}, 2673 (2010).

\bibitem{danneau2008}
R.~Danneau, F.~Wu, M.~F.~Craciun, S.~Russo, M.~Y.~Tomi, J.~Salmilehto,
A.~F.~Morpurgo, and P.~J.~Hakonen, 
Phys. Rev. Lett. {\bf 100}, 196802 (2008).

\bibitem{balandin2013}
A.~A.~Balandin, 
Nature Nanotech. {\bf 8}, 549 (2013).

\bibitem{wurm2009}
J.~Wurm, M.~Wimmer, I.~Adagideli, K.~Richter, and H.~U.~Baranger,
New J. Phys. {\bf 11}, 095022 (2009).

\bibitem{li2009}
H.~Li, L.~Wang, Z.~Lan, and Y.~Zheng,
Phys. Rev. B {\bf 79}, 155429 (2009).

\bibitem{zhou2010}
B.~Zhou, B.~Zhou, W.~Liao, and G.~Zhou,
Phys. Lett. A {\bf 374}, 761 (2010).

\bibitem{castroneto2009} 
A.~H.~Castro Neto, F.~Guinea, N.~M.~R.~Peres, K.~S.~Novoselov, and A.~K.~Geim,
Rev. Mod. Phys. {\bf 81}, 109 (2009).

\bibitem{brey2006}
L.~Brey and H.~A.~Fertig,
Phys. Rev. B {\bf 73}, 235411 (2006).

\bibitem{kp2011}
P.~Marconcini, M.~Macucci,
La Rivista del Nuovo Cimento {\bf 34}, 489 (2011), DOI: 10.1393/ncr/i2011-10068-1.

\bibitem{Bardarson}
J.~H.~Bardarson, J.~Tworzyd{\l}o, P.~W.~Brouwer, and C.~W.~J.~Beenakker,
Phys. Rev. Lett. {\bf 99}, 106801 (2007).

\bibitem{Nomura}
K.~Nomura, M.~Koshino, and S.~Ryu,
Phys. Rev. Lett. {\bf 99}, 146806 (2007).

\bibitem{tworzydlo2008}
J.~Tworzyd{\l}o, C.~W.~Groth, and C.~W.~J.~Beenakker,
Phys. Rev. B {\bf 78}, 235438 (2008).

\bibitem{stacey1982}
R.~Stacey, 
Phys. Rev. D {\bf 26}, 468 (1982).

\bibitem{bender1983}
C.~M.~Bender, K.~A.~Milton, and D.~H.~Sharp,
Phys. Rev. Lett. {\bf 51}, 1815 (1983).

\bibitem{hernandez2012}
A.~R.~Hern\'andez and C.~H.~Lewenkopf,
Phys. Rev. B {\bf 86}, 155439 (2012).

\bibitem{susskind1977}
L.~Susskind, 
Phys. Rev. D {\bf 16}, 3031 (1977).

\bibitem{snyman2008}
I.~Snyman, J.~Tworzyd{\l}o, and C.~W.~J.~Beenakker,
Phys. Rev. B {\bf 78}, 045118 (2008).

\bibitem{chalker1988}
J.~T.~Chalker and P.~D.~Coddington,
J. Phys. C {\bf 21}, 2665 (1988).

\bibitem{ho1996}
C.-M.~Ho and J.~T.~Chalker,
Phys. Rev. B {\bf 54}, 8708 (1996).

\bibitem{connolly2012}
M.~R.~Connolly, R.~K.~Puddy, D.~Logoteta, P.~Marconcini, M.~Roy,
J.~P.~Griffiths, G.~A.~C.~Jones, P.~A.~Maksym, M.~Macucci, and C.~G.~Smith,
Nano Lett. {\bf 12}, 5448 (2012).

\bibitem{logoteta2012}
D.~Logoteta, P.~Marconcini, M.~R.~Connolly, C.~G.~Smith, and M.~Macucci,
in Proceedings of IWCE 2012, IEEE Conference Proceedings, 6242841 (2012),
DOI: 10.1109/IWCE.2012.6242841.

\bibitem{CCnote}
The Chalker-Coddington model of Ref.~\cite{snyman2008} describes just one valley of the graphene spectrum, 
so it is impossible to implement in it the physical boundary conditions.

\bibitem{PT}
M.~Fagotti, C.~Bonati, D.~Logoteta, P.~Marconcini, and M.~Macucci,
Phys. Rev. B {\bf 83}, 241406(R) (2011).

\bibitem{ando2005}
T.~Ando, 
J. Phys. Soc. Jpn. {\bf 74}, 777 (2005).

\bibitem{rothe}
H.~J.~Rothe, {\em Lattice Gauge Theories: An Introduction}, 
World Scientific Lecture Notes in Physics 74
(World Scientific, Singapore, 2005).

\bibitem{MM}
I.~Montvay and G.~M\"unster,
{\em Quantum Fields on a Lattice}
(Cambridge University Press, Cambridge, 1994).

\bibitem{smilga}
A.~Smilga,
{\em Lectures on Quantum Chromodynamics}
(World Scientific, Singapore, 2001).

\bibitem{BF:2013}
C.~Bonati and M.~Fagotti (unpublished).

\bibitem{korner1988}
T.~W.~K\"orner, {\em Fourier Analysis}
(Cambridge University Press, Cambridge, 1988).

\bibitem{wolff2012}
J.~F\"orster, A.~Saenz, and U.~Wolff,
Phys. Rev. E {\bf 86}, 016701 (2012). 

\bibitem{datta} S.~Datta, {\em Electronic Transport in Mesoscopic Systems}
(Cambridge University Press, Cambridge, 1995).

\bibitem{vides1} http://vides.nanotcad.com

\bibitem{vides2}
G.~Fiori and G.~Iannaccone,
Proc. IEEE {\bf 101}, 1653 (2013).

\bibitem{ashcroft}
N.~W.~Ashcroft and N.~D.~Mermin, {\em Solid State Physics}
(Thomson Learning, London, 1976), p.~341.

\bibitem{saito}
R.~Saito, G.~Dresselhaus, and M.~S.~Dresselhaus, 
{\em Physical Properties of Carbon Nanotubes}
(Imperial College Press, London, 1998).


\end{thebibliography}
\end{document}